\documentclass[%
 reprint,
superscriptaddress,
 amsmath,amssymb,
 aps,
]{revtex4-1}
\usepackage{color}
\usepackage{xcolor}
\usepackage{amsthm}
\usepackage{graphicx}
\usepackage{dcolumn}
\usepackage{bm}
\usepackage{mathrsfs}
\usepackage{dsfont}
\usepackage{ulem}
\usepackage{selinput}
\usepackage{mathtools}

\begin{document}

\preprint{APS/123-ART}

\title{Quantum Populations near Black-Hole Singularities}  

\author{Ludwig Eglseer}%
 \email{l.eglseer@physik.uni-muenchen.de}
 \affiliation{Arnold Sommerfeld Center for Theoretical Physics, Theresienstra{\ss}e 37, 80333 M\"unchen, Germany\\}
 \author{Stefan Hofmann}%
 \email{stefan.hofmann@physik.uni-muenchen.de}
\affiliation{Arnold Sommerfeld Center for Theoretical Physics, Theresienstra{\ss}e 37, 80333 M\"unchen, Germany\\}
  \author{Marc Schneider}%
 \email{marc.schneider@aei.mpg.de}
 \affiliation{Albert-Einstein Institute for Gravitational Physics, Am M\"uhlenberg 1, 14476 Potsdam, 
 Germany\\}

\date{\today}

\begin{abstract}
Schwarzschild black-hole interiors border on space-like singularities representing 
classical information leaks. 
We show that local quantum physics is decoupled from these leaks due to 
dynamically generated boundaries, called Zeno borders. 
Beyond Zeno borders black-hole interiors become asymptotically silent, 
and quantum fields evolve freely towards the geodesic singularity with vanishing
probability measure for populating the geodesic boundary. 
Thus Zeno borders represent a probabilistic completion of Schwarzschild 
black holes within the semiclassical framework.
 \end{abstract}

 \keywords{Suggested keywords}
 \maketitle
\section{Introduction}
Schwarzschild black-hole interiors contain geodesic borders separating them 
from space-like singularities. Any information migrating across a geodesic border
towards the singularity is irretrievably lost by causality. 
If Schwarzschild singularities absorb information, the corresponding evolution 
qualifies as paradoxical since it violates sacrosanct rules of information processing \cite{hor04}. 
It is generally expected that spacetime fluctuations deform the Schwarzschild geometry 
near its geodesic border to yield a consistent quantum evolution. 
While the details of this dynamical regularization mechanism are unknown, they 
are important for global aspects of quantum information processing by black holes,
such as the black-hole information paradox \cite{mat09, gid12,ma17}. 

In this article, we show that Schwarzschild singularities border on asymptotically
silent spacetime regions, that is regions inhibiting spatial quantum correlations 
irrespective of the initial field configuration. 
More importantly, they accommodate so-called Zeno borders, which 
mark a stack of hypersurfaces terminated by a geodesic border 
with the following property: The probability measure for populating quantum information 
within the stack decreases monotonously towards the singularity and vanishes 
at the geodesic border. 
As a consequence, quantum events cannot probe the geodesic border and 
quantum information cannot migrate across the geodesic border, irrespective of any
quantum completion of gravity.
In other words, Zeno borders render black-hole interiors leak-proof and 
represent, to the best or our knowledge, the first explicit quantum completion 
of a geodesically incomplete spacetime within the semiclassical framework 
\footnote{A probabilistic completion as presented here, 
has been formerly described in quantum mechanics, e.g. for timelike singularities \cite{hor90} }.
Quantum field theoretic 
completeness conceptualises the intrinsic consistency of quantum
field theory in curved space-times in the sense that the theory itself is protected from a 
breakdown when reaching the geodesic border. Zeno regions incorporate this dynamical 
protection mechanism which guarantees the predictability of the theory itself and
likewise passes the regularity on to physical observables.

To substantiate this probabilistic completion we proceed as follows: 
after stating the geometrical set-up, we will present an intuitive argument based on 
scaling relations in the microp{\ae}dia which introduces the idea of the Zeno region from the 
perspective of quantum measurements. The exact argument is then given in the macrop{\ae}dia
followed by an interpretation based on observables showing that the probabilistic completion 
closes the geodesic border with respect to quantum field theory. 


\section{Geometrical Preliminaries}
Schwarzschild black holes are the warped geometries 
$\mathcal{B}:=P_<\times_t S^2$, with $P_<$ denoting the 
region 
$t<t_{\rm g}:= 2 M$ in the $(t,r)$-half plane $(0, r_{\rm g})\times\mathbb{R}^+$,
where the projection $t:\mathcal{B}\rightarrow (0,r_{\rm g})$ is the 
Schwarzschild time and the projection
$r:\mathcal{B}\rightarrow \mathbb{R}^+$ is the Schwarzschild
radius function, 
and $S^2$ denotes the unit two-sphere.
Note that in these conventions the Schwarzschild coordinate vector field 
$\partial_t$ is time-like, and $\partial_r$ is space-like on $\mathcal{B}$.
Compared to the Schwarzschild exterior spacetime, Schwarzschild time 
and radius function have interchanged their meaning. 
Taking this into account, the quadratic form associated with 
Schwarzschild black holes is 
${\rm d}s^2=-s^{-1}(t){\rm d}t^2 + s(t) {\rm d}r^2 + t^2 w$,
where $s(t):=|1- r_{\rm g}/t|$ is the Schwarzschild function, and $w$ denotes 
the line element associated with the Euclidean metric on $S^2$, equipped 
with the usual spherical coordinates $(\vartheta,\varphi)$.
$\mathcal{B}$ corresponds to a spatially homogeneous and anisotropic 
cosmological solution of general relativity. 
It can be foliated into
Cauchy hypersurfaces $(0,r_{\rm g})\times \Sigma$ along the temporal direction,
where $\Sigma$ is 
the folio of spatial hypersurfaces $\Sigma_t$ labeled by Schwarzschild time.
The metric field associated with the above quadratic form will be denoted by $g$,
and its pull-back to $\Sigma$ by $g_{_\Sigma}$.

In this geometry, the geodesic motion of a point particle that is initially equatorial 
relative to Schwarzschild spherical coordinates is bound to remain equatorial,
$\vartheta=\pi/2$. The so-called energy equation
$E^2=({\rm d}t/{\rm d}s)^2+V_{\rm eff}$ holds, where
$E:=s(t){\rm d}r/{\rm d}s$ and $L:=t^2 {\rm d}\varphi/{\rm d}s$ are constants,
which have an intuitive interpretation in the exterior as asymptotic energy per unit mass 
and angular momentum per unit mass, respectively. In fact, in the exterior, the definition 
of $L$ formally coincides with Kepler's second law. The effective potential 
is given by $V_{\rm eff}:=-(1+L^2/t^2)s(t)$. 
Close to the endpoint at $t=0$, the effective potential is bounded from above,
$V_{\rm eff}=-L^2 r_{\rm g}/t^3$ plus less singular contributions. 
Thus the classical motion generated by $V_{\rm eff}$ is incomplete at $t=0$.
Potential incompleteness implies that $\mathcal{B}$ is geodesic incomplete \cite{hawk70},
which in turn qualifies $\Sigma_{0}$ as a geodesic information sink. 
Geodesic incompleteness, however, does not imply quantum incompleteness (and vice versa). 

\section{Microp{\ae}dia}
Let us first give an intuitive argument based on scaling relations
showing that quantum information in $\mathcal{B}$ cannot migrate across $\Sigma_{0^+}$, 
before providing precise statements.
In $\mathcal{B}$, 
scalar quantum fields $\Phi^*, \Phi$ charged under $U(1)$ 
evolve according to the Lagrange density
$\mathcal{L}= \mathcal{L}_0+\mathcal{L}_{\rm int}$, with the first term denoting 
the free theory $\mathcal{L}_0 = \Phi^* \Box \Phi$. The intuitive argument
will be given in the absence of interactions, $\mathcal{L}_{\rm int}\equiv 0$.
Close to the space-like singularity $\Sigma_0$ bordering on $\mathcal{B}$,
${\rm d}s^2\cong -(t/r_{\rm g}) {\rm d}t^2+(r_{\rm g}/t) {\rm d}r^2 + t^2 w$,
where $\cong$ means equality up to sub-leading contributions in each term 
as $\Sigma_0$ is approached \footnote{For $\vartheta $ small, the quadratic form can be
transformed into type-D Kasner line-element with exponents 
$(p_1,p_2,p_3)=(2/3,2/3,-1/3)$, corresponding to a spatially anisotropic
cosmology. }.
In this asymptotic regime, 
$\Box \cong  (-r_{\rm g}/t) (\partial_t^{\; 2} + (1/t) \partial_t)+
 (t/r_{\rm g}) \partial_r^{\; 2} + (1/t^2)\partial_\sphericalangle^{\; 2}$.
Here $\partial_\sphericalangle^{\; 2}$ denotes the usual angular part 
of the Laplace operator in $\mathbb{R}^3$ in Schwarzschild spherical coordinates.

The corresponding Green function $G$ is sourced by 
$\delta(t-t^\prime)\delta(\sigma-\sigma^\prime)/\sqrt{-{\rm det}(g)}$, with 
$\sigma$ and $\sigma^\prime$ denoting Schwarzschild spatial
coordinates of events localized on $\Sigma_t$ and $\Sigma_{t^\prime}$, respectively, 
and satisfies
\begin{eqnarray}
	&&D(t) G(t,\sigma ; t^\prime, \sigma^\prime)
	=
	\delta(t-t^\prime)\delta(\sigma-\sigma^\prime)
	\; ,
	\nonumber \\
	&&D(t) :=
	-r_{\rm g} \partial_t (t\partial_t) 
	+
	(t/r_{\rm g}) t^2 \partial_r^{\; 2} 
	+ \partial_\sphericalangle^{\; 2} 
	\; . \label{kgg}
\end{eqnarray}
In order to estimate the asymptotic relevance of each term in the 
differential operator, consider $D(\varepsilon \tau)$ in the limit 
$\varepsilon \rightarrow 0^+$. 
The effective potential for free fields scales asymptotically as
$1/t^2$ and develops a repulsive barrier. 
It is well-known from quantum mechanics
that potentials of this type cannot be penetrated via tunneling processes.

Following the geometrical 
description of space-like singularities by 
Belinskii, Khalatnikov, and Lifshitz \cite{bkl70}, temporal variations dominate 
over spatial variations in the region bordering on $\Sigma_{0^+}$.
Therefore,
 $D(\varepsilon\tau)\cong (1/\varepsilon) \partial_\tau(\tau\partial_\tau)$.
 The time-dependent part of the source distribution scales like $
\delta(\varepsilon\tau-t')$.
 This effectively allows to split the Green function 
 $G=T(t,t^\prime) P(\sigma,\sigma^\prime)$ in the vicinity of $\Sigma_{0^+}$, 
 with the asymptotic dynamics given by $\partial_\tau(\tau\partial_\tau T) \cong 0$.
 Here, all identifiers labeling the eigenvalue problem of the Laplace operator 
 have been suppressed for ease of notation. We find the asymptotic solution
 $T(t,t^\prime) \cong C_0(t^\prime)+
  C_1(t^\prime) \ln(t/r_{\rm g})$, with 
 $C_{0,1}<\infty$ depending on the initial values given
 on $\Sigma_{t^\prime}$.
 
\begin{figure}
\includegraphics[width=8.6cm]{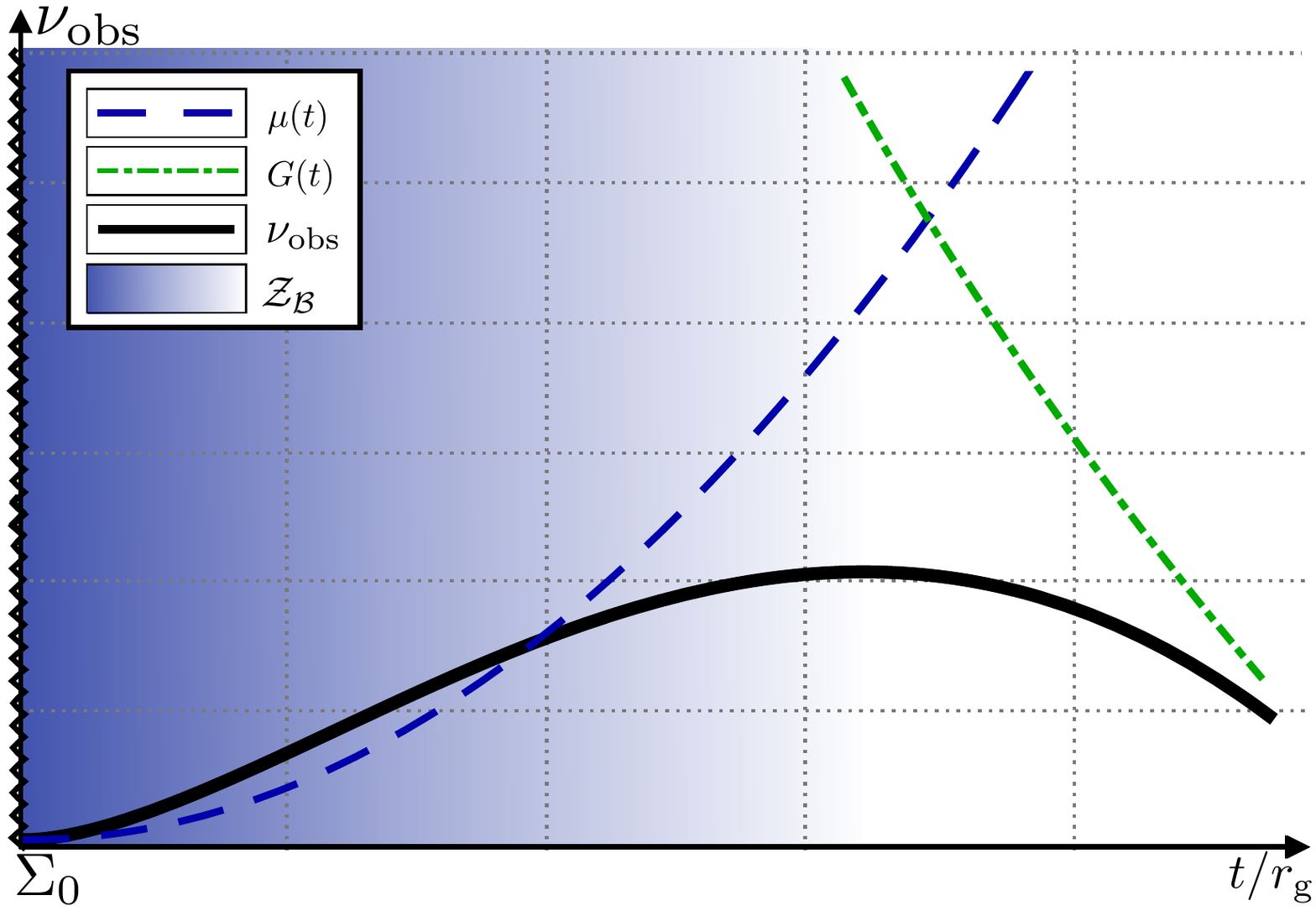}
\caption{
\label{fig:plot1} 
Vertex density $\nu_{\rm obs}$ for measurement processes inside Schwarzschild black holes.
The communication between emitter and absorber, described by the asymptotic Green function 
$G$, diverges towards the geodesic border $\Sigma_0$, but the integration measure $\mu$
goes sufficiently fast to zero to render $\nu_{\rm obs}$ finite at all times. 
Furthermore, within the Zeno region (shaded), see Fig.~\ref{fig:plot2} and the 
{\it macrop\ae dia} for details, $\nu_{\rm obs}$ is monotonously decreasing 
and vanishes towards $\Sigma_0$.
}
\end{figure}

In order to appreciate the rather mild divergence of the asymptotic solution $T$,
we introduce a model for a physical detection with an emitter $F_{\rm em}$ localized on 
$\Sigma_{t^\prime}\;,  t^\prime\in (0,t_*)$ in the asymptotic domain, where $t_*$ 
denotes a fiducial time in this regime, and
an absorber $F_{\rm ab}$ on $\Sigma_{\varepsilon\tau}$.
A general measurement is given by a vertex density and involves the communication
channel between source and the detector.
For instance, consider $F_{\rm ab} = \delta(t-\varepsilon \tau) \; f_{\rm ab} (\sigma)$,
with $f_{\rm ab}$ encoding the spatial extension of the detector on 
$\Sigma_{\varepsilon\tau}$. This blueprint effectively replaces part of $\Sigma_{0^+}$
with a detector volume that can resolve arbitrary frequencies. 
Note that the asymptotic regime is controlled by the parameter $\varepsilon$ while
$\tau$ represents a constant instant of time.
The classical measurement process is described by the on-shell vertex density 
$\nu_{\rm obs}= \sqrt{{-\rm det}(g)} F_{\rm ab}^{\; *} \Xi_{\rm os} + {\rm c.c.}$
with $\Xi_{\rm os}=G\ast F_{\rm em}$ denoting a convolution of $F_{\rm em}$ 
with a bi-local kernel given by $G$ with respect to $x'$ \cite{dewitt03}. 
In the region bordering on $\Sigma_{0^+}$, as specified 
by the support properties of emitter and absorber, 
$\nu_{\rm obs} \cong 
t^2 {\rm ln}(t) \delta(t-\varepsilon\tau) {\rm sin}(\vartheta) f_{\rm ab} \; F_{\rm em}$,
where $F_{\rm em}$ contains the exclusive information on the emission process and 
depends only on source parameters. In particular, $F_{\rm em}$ is finite in accordance
with the homogeneous interior $\mathcal{B}$. For fixed $t'$ and
$\varepsilon\rightarrow 0$, the measurement of the emitter's influence on the detector 
gives a vanishing response, see figure \ref{fig:plot1},
$\nu_{\rm obs} \cong 0$, in the distributional sense. 
This implies that no information carried by local bookkeeping devices
reach $\Sigma_{0^+}$.

It is possible to be more specific about the emitter and absorber. As an example, the energy momentum tensor for the complex scalar field
scales like 
$\mathcal{T}=$d$\Phi \otimes $d$\Phi^* \propto 1/(\varepsilon \tau)^2$ on 
$\Sigma_{\varepsilon \tau}$ and develops a singularity towards $\Sigma_{0^+}$.
Our detector model is given by
$F_{\rm ab} = \delta(t-\varepsilon \tau) \; f_{\rm ab} (\sigma)$,
with $f_{\rm ab}$ encoding the spatial extension of the detector on 
$\Sigma_{\varepsilon\tau}$ or a more physical absorber of an ideal gas
$\widetilde{F}_{\rm ab}=M(\varepsilon\tau) U\otimes U$, where 
$U\cong \sqrt{\varepsilon\tau/r_{\rm g}}{\rm d}t$ is 
the normal, and $M(t)$ 
denotes the spatial volume integral over an energy density.
From a phenomenological point of view, $F_{\rm ab}$ is required 
to have nontrivial support towards $\Sigma_{0^+}$. 
Then, $\widetilde{\nu}_{\rm obs}\cong 0$, as well, which only confirms that
the asymptotic description of the tree-level measurement process 
is independent of the tensor providing the principal communication channel.

Before closing the microp\ae dia, let us briefly discuss 
the asymptotic diagnostics of Noether charges. 
Let $\Phi:\mathcal{B}\rightarrow\mathbb{C}$ carry a $U(1)$ charge.
The four-dimensional $U(1)$ current density is $j=i(\Phi^* P \Phi - c.c.)$, where 
$P$ denotes the four-momentum operator. Projecting the current density 
onto $U$ from before, we find the following scaling relation for the charge 
density $\rho$ localized  on $\Sigma_{\varepsilon\tau}$:
$\rho(\varepsilon\tau)\cong \rho(t_*) (t_*/\varepsilon\tau)^{3/2}$,
which formally diverges as $\Sigma_{0^+}$ is approached. 
Physical measurements
of the charge $Q(\varepsilon\tau)$, however, yield a finite result. In fact 
$Q(\varepsilon\tau)=Q(t_*)$. Black holes cannot be discharged 
through the geodesic singularity bordering on $\Sigma_{0^+}$. 
Any active information sink would necessarily lead to charge depletion. 
Note that this discussion of asymptotic charge conservation is fully based
on local physics inside black holes, and no reference to the usual global 
characterization in the exterior is made. 

\section{Macrop{\ae}dia}
A more rigorous argument is based on the Schr\"odinger representation
of local quantum physics which we summarise for convenience below. 
Let $(a,b)\times\Sigma$ be a Cauchy foliation of a globally hyperbolic spacetime $(\mathcal{M},g)$
with a geodesic border $\Sigma_a$ in the folio $\Sigma$ of hypersurfaces. 
For $t\in(a,b)$, we denote by $\mathcal{C}(\Sigma_t)$ the space of instantaneous 
field configurations $\phi:\Sigma_t\rightarrow\mathbb{C}$. 
Consider the $\mathbb{C}-$vector space 
of measurable wave functionals $\Psi_t: \mathcal{C}(\Sigma_t)\rightarrow \mathbb{C}$ 
whose modulus is square integrable with respect to the formal functional measure $\mathcal{D} \phi$. 
Let $L^2(\mathcal{C}(\Sigma_t), \mathcal{D}\phi)$ be
the quotient of this space by the subspace of wave functionals vanishing $\mathcal{D}\phi-$almost
everywhere in $\mathcal{C}(\Sigma_t)$. We introduce the usual norm on
$L^2(\mathcal{C}(\Sigma_t), \mathcal{D}\phi)$. If $\mathcal{U}$ is a formally measurable 
subset of $\mathcal{C}(\Sigma_t)$ and $\mathcal{X}_\mathcal{U}$ the corresponding indicator functional,
then $\|\mathcal{X}_\mathcal{U}{\Psi_t}\|^2$ is the probability for the field configuration on $\Sigma_t$ 
to be given by some $\phi\in\mathcal{U}$.
This interpretation requires normalizable wave functionals. 
The configuration field operator $\Phi[f]$ is just the multiplication operator with $\phi[f]$, 
where $f$ denotes some smooth test function of compact support on $\Sigma_t$ such that 
$\langle\Psi_t| \Phi[f] |\Psi_t\rangle=\|\sqrt{\phi[f]} \Psi_t\|^2$ is bounded.  
The conjugated momentum field operator $\Pi[f]$ is the functional derivative 
$-\mathrm{i} ({\rm det}(g_{\Sigma_t}))^{-1/2} \delta/\delta \phi$ in the direction of $f$. 
Heisenberg's fundamental uncertainty relation is a consequence of 
$[\Phi[f_1],\Pi[f_2]]=\mathrm{i} (f_1,f_2)$, where $(\cdot,\cdot)$ denotes the 
scalar product of smearing functions on $\Sigma_t$.

Consider the evolution semigroup given by the continuous map 
$(a,b) \rightarrow L^2(\mathcal{C}(\Sigma_t), \mathcal{D}\phi)$, defined 
by $t\rightarrow \mathcal{E}(t,t_0) |\Psi_{t_0}\rangle$ for an initial time $t_0\in (a,b)$. 
A probabilistic interpretation is only possible for $\|\mathcal{E}(t,t_0)\|\le 1$, that is 
for contractive evolution semigroups. While this includes unitarity as a special case 
for Minkowski spacetime, its generator density $\mathcal{H}$ is only required to be 
accretive, ${\rm Re}(\langle\Psi_t|\mathcal{H}|\Psi_t\rangle)\ge 0$, instead of being
self-adjoint. On generic spacetimes, the pair (contractive, accretive) supersedes (unitary, self-adjoint) 
as a requirement on evolution semigroups in order to allow for a probabilistic interpretation. 
This has important ramifications for physical spacetimes as support for quantum events:

Let $(L^2(\mathcal{C}(\Sigma_{t_0})), \mathcal{H})$ be the initial Schr\"odinger representation
of a quantum field theory in a globally hyperbolic  spacetime $(M,g)$ with the Cauchy foliation 
$(a,b)\times\Sigma$ containing a geodesic boundary $\Sigma_a$ in the folio $\Sigma$
of hypersurfaces. A Zeno region $\mathcal{Z}_\mathcal{M}$ in $\mathcal{M}$ 
is a stack of Cauchy hypersurfaces in $\Sigma$ with the following property:
The evolution semigroup generated by $\mathcal{H}$ 
is strictly contractive in $\mathcal{Z}_\mathcal{M}$ and $\|\mathcal{E}(c,t_0)\|\rightarrow 0$ 
as $c\rightarrow a$. 

Evidently, $\mathcal{Z}_\mathcal{M}$ is bounded by some Cauchy hypersurface $\Sigma_z$,
called the Zeno border, and the geodesic border $\Sigma_a$. Let us justify the introduction of 
Zeno regions: If a spacelike singularity borders on a Zeno region, then the singularity
cannot be probed by quantum fields and the quantum evolution in its immediate vicinity 
allows for a probabilistic interpretation. This distinguishes the physical spacetime as the principal support
of quantum events from the geometrical spacetime model. Therefore Zeno regions imply 
a probabilistic or quantum completion of a geodesic incomplete spacetime within the
semiclassical framework. 

Let us now turn back to Schwarzschild black-hole interiors $\mathcal{B}$, 
Cauchy-foliated as $(0, r_{\rm g})\times \Sigma$ with respect to the Schwarzschild time function. 
Consider the space $\mathcal{C}(\Sigma_{t_0})$ of instantaneous 
field configuration $\phi,\phi^*:\Sigma_{t_0}\rightarrow \mathbb{C}$ and its 
Schr\"odinger representation $(L^2(\mathcal{C}(\Sigma_{t_0})), \mathcal{H})$ in $\mathcal{B}$
with Hamilton density $\mathcal{H}=\mathcal{G}(\Pi^*,\Pi)+ \mathcal{V}(\Phi^*,\Phi)$, 
where $\mathcal{G}(\Pi^*,\Pi):=\sqrt{-g_{tt}} \, \Pi^*\Pi/{\rm det}(g_{_{\Sigma_t}})$, and
$\Pi:=-{\rm i}\delta/\delta \Phi$ denotes the momentum field conjugated to $\Phi$ \cite{kiefer}.
The effective potential density 
$\mathcal{V} := \sqrt{-g_{tt}}(|{\rm grad}\Phi|^2+m^2|\Phi|^2)$
is a pure multiplication operator, with $m$ being the mass of the scalar field.

For the ground-state wave functional we make the following ansatz 
\begin{eqnarray}
\label{Gauss}	
	\Psi_t[\phi^*,\phi]
	=
	\mathcal{N}_t
	\exp\left(-[V^*]\mathcal{K}_t[V]\right)
	\; ,
\end{eqnarray}	
which will be justified a posteriori. 
Here, $V=(\phi,\phi^*)^{\rm T}$, and 
\begin{eqnarray}
\label{K}
	[V^*]\mathcal{K}_t[V]
	:=
	\frac{1}{2}
	\int_{\Sigma_t} {\rm d}\mu_{x,y} \;
	V^*(x) K_t(x,y) V(y)
\end{eqnarray} 
is a quadratic functional with a bi-local representation
given by the matrix $K$. In the absence of interactions,
$K$ is diagonal and only its trace enters in (\ref{K}).
Close to the Schwarzschild singularity $\Sigma_0$
the quantum evolution trivialises \cite{hof17}, that is $\mathcal{H}\approx \mathcal{G}(\Pi^*,\Pi)$.
As a consequence, the trace of the bi-local kernel matrix $K$ becomes a
contact term on $\Sigma_t\in \Sigma$ near the geodesic border, 
${\rm tr}(K_{\varepsilon\tau})(x,y)\cong k(\varepsilon\tau) \delta^{(3)}(x,y)$ 
with \cite{hof15}
\begin{eqnarray}
\label{resold}
	{\rm Im}\left(k(\varepsilon\tau)\right)
	&\cong &
	- \frac{2}{(\varepsilon \tau)^3} \frac{1}{|{\rm ln}(\varepsilon\tau)|}
	\; , 
	\nonumber \\
	{\rm Re}\left(k(\varepsilon\tau)\right)
	&\cong &
	\left|{\rm Im}(C)\right|
	\frac{\left|{\rm Im}\left(k(\varepsilon\tau)\right)\right|}
	{\left|{\rm ln}(\varepsilon\tau)\right|}
	\; ,
\end{eqnarray}
where $\cong$ refers to the limit $\varepsilon\rightarrow 0$.
Our result is reminiscent of the analysis of
Kasner spacetimes by Belinskii, Khalatnikov and Lifshitz:
In the vicinity of space-like singularities, but still in the 
domain of general relativity, the spatial variations of 
local quantities are insignificant compared to temporal gradients. 
This is also similar to asymptotic silence \cite{an05} which describes 
how the particle horizon shrinks in homogeneous but anisotropic spacetimes 
bordering on a spacelike singularity. All spatial correlations vanish eventually. 

\begin{figure}
\includegraphics[width=8.6cm]{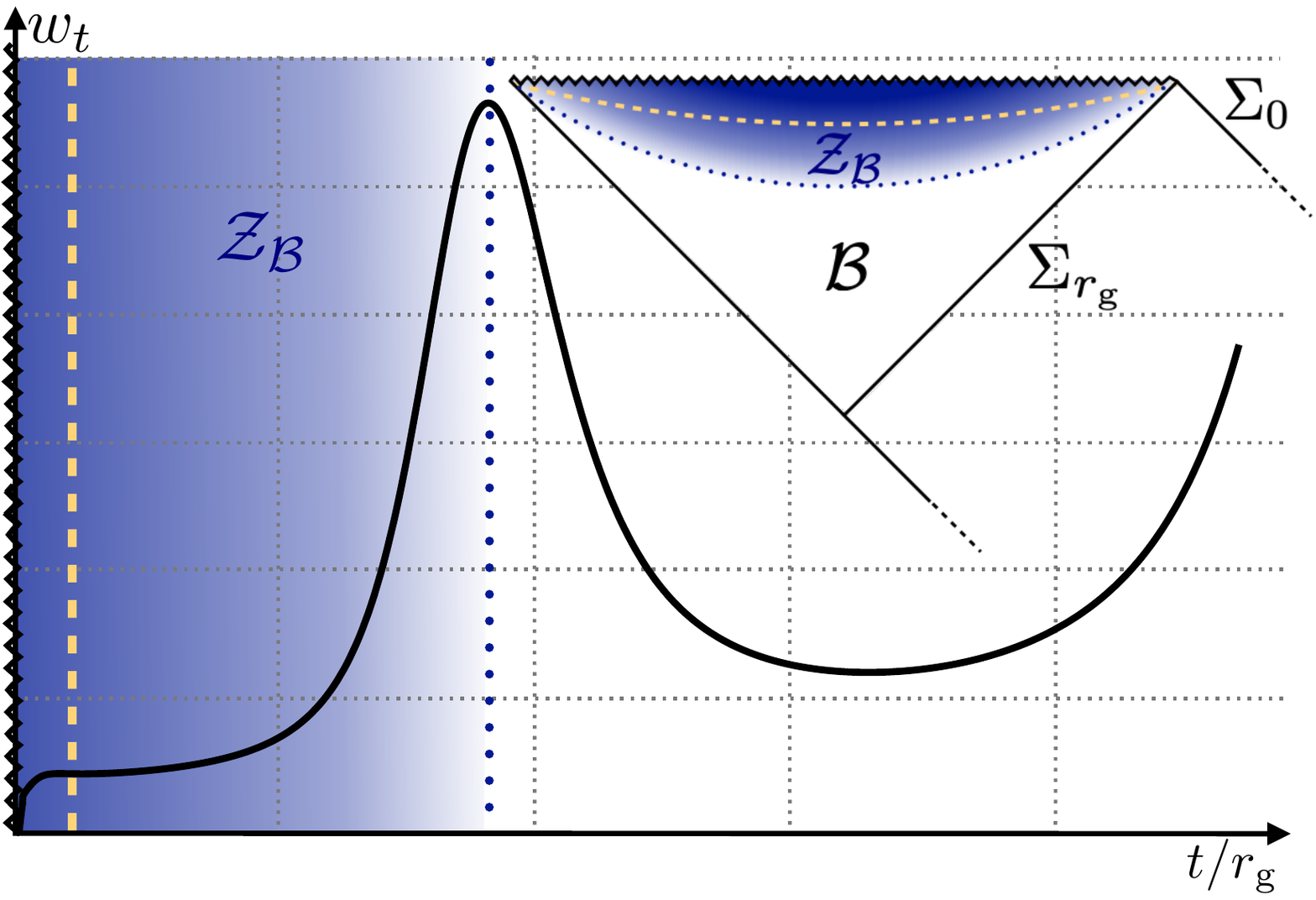}
\caption{\label{fig:plot2} Probability density $w_t^\ell=|\Psi_t^\ell[\phi^*,\phi]|^2$ for the distribution of a field configuration $\phi$ for a fixed $\ell$
in the folium $\Sigma$ of $\mathcal{B}$. Wavy lines mark the geodesic border $\Sigma_0$, dashed lines represent the 
onset of asymptotic silence, and the shaded region between $\Sigma_0$ and the dotted line is the Zeno region
$\mathcal{Z}_\mathcal{B}$. The upper right corner displays the corresponding Penrose diagram.}
\end{figure}

The asymptotic kernel (\ref{resold}) implies that the immediate vicinity 
of the geodesic border is a Zeno region $\mathcal{Z}_\mathcal{B}$ since
\begin{eqnarray}\label{phi2}
	\lim\limits_{\varepsilon\rightarrow 0} \Psi_{\varepsilon\tau}[\phi^*,\phi]
	=
	\lim\limits_{\varepsilon\rightarrow 0}
	\left|\ln\left(\varepsilon\tau\right)\right|^{-\Lambda v(\Sigma_{\varepsilon\tau})}
	\; =0,
\end{eqnarray}
where $\Lambda$ denotes a short-distance cut-off and 
$v(\Sigma_{\varepsilon\tau})$ a volume regularization. 
Note that any hypersurface contained in $\mathcal{Z}_\mathcal{B}$
has an infinite volume, 
\eqref{phi2} is robust 
against the detailed regularization prescription. 
The Zeno region $\mathcal{Z}_\mathcal{B}$ secludes the Schwarzschild black-hole's interior
 not granting any 
sojourn to quantum probes on the singular hypersurface bordering on $\mathcal{B}$. Thence
the physical space-time is closed for quantum probes despite being geodesically incomplete in the 
mathematical notion.
In other words, it is the portion of $\mathcal{B}$ that can be probed via the support of quantum events, see 
Fig. \ref{fig:plot2}.  

Intuitively $\mathcal{Z}_\mathcal{B}$ marks the
stack of hypersurfaces which enforces the probabilistic completion such that the geodesic border
is closed in the quantum theoretical sector. 
Quantum field theory, hence, introduces a new scale 
$\tau_\mathcal{Z}$ in the black hole interior.
To classify $\tau_\mathcal{Z}$ further, 
we compare it to the scales coming from classical gravity, i.e. $r_{\rm g}$, and 
from quantum gravity, i.e. the Planck time $t_{\rm Pl}$.


Following the definition of $\mathcal{Z}_\mathcal{B}$, the exact location is given
by an interplay between the background geometry and the probing quanta.
The relevant quantity to estimate $\tau_\mathcal{Z}$
is the position of the maximum in the probability density $w_{\varepsilon \tau}=
|\Psi_{\varepsilon \tau}[\phi^*,\phi]|^2$ which is closest to
$\Sigma_0$. Since the asymptotic solution is confined to a small region, we need to 
extend it by including less singular terms of order 
$\mathcal{O}(1/\varepsilon)$ such that $\Psi_{\varepsilon\tau}$ covers the evolution of test fields
in a larger portion of $\mathcal{B}$ and a possible maximum becomes visible. In this range, we find
oscillatory solutions to \eqref{kgg} given by Bessel functions. This allows to estimate
the position of $\tau_\mathcal{Z}$ at which the maximum occurs which also
determines the beginning of  
the Zeno region. To give a proof of concept, we perform a mode-sum decomposition in $l\in\mathbb{N}_0$, where we split off the time-dependent 
contribution of the kernel (note, for the full kernel  the sum over all $l$ has to be performed) 
\begin{equation}
k_\ell(\varepsilon \tau)=\frac{-i}{\varepsilon^3 \tau^2}\partial_{\tau}\ln
\left(c_1{\rm J}_0\left(\sqrt{\ell^2\varepsilon\tau}\right)+c_2{\rm Y}_0\left(
\sqrt{\ell^2\varepsilon\tau}\right)\right)\;.
\end{equation}
Here, $\ell^2=4l(l+1)$, $c_i\in\mathbb{C}$ constants of integration,
and J$_0$ and Y$_0$ are the Bessel function of the first and second kind. The location of
$\Sigma_\mathcal{Z}^\ell$ can be expressed as a
three-parameter family of hypersurfaces determined 
by $\ell^2$ as well as the initial conditions represented by $c_1$ and $c_2$ of the quantum probes. By performing a series expansion for small 
arguments $\ell^2\varepsilon\tau\ll1$, we can approximate the position of the maximum in $w_\tau^\ell$ 
for $\ell^2$-values that allow for a suitable covering in the $\tau$-direction within this 
approximation. Thus, we find for a given $\ell$
\begin{eqnarray}
\label{maxZeno}	
 \tau_\mathcal{Z}^\ell
 \approx 
 \frac {r_{\rm g}}{|\mathcal{C}(c_1,c_2)|^2\ell^2}
	\; ,
\end{eqnarray}	
where $\mathcal{C}(c_1,c_2)$ is a constant depending on the initial conditions. 
Apparently, increasing the angular momentum $\ell$ results in shifting the Zeno border closer 
to the geodesic border while
for the case $\ell=0$ we find the logarithmic (non-oscillating) solution in \eqref{phi2} 
where, the Zeno border lies at infinity, i.e. it covers all of the interior.
Although the exact position of 
$\mathcal{Z}_\mathcal{B}$ depends on the specifics of the solution, we can compare $\tau_\mathcal{Z}^\ell$ to 
the high-energy scale provided by the Planck energy. To this aim, we consider Christensen's general form of
a renormalised stress-energy tensor \cite{chri76} 
with Synge's world function given in \cite{Levi15} and look at its trace
$\langle T(x)\rangle_{\rm ren}=\tfrac{1}{8\pi^2}\sqrt{-{\rm det}(g)}\left(\tfrac{1}{360}K + C\right)+\mathcal{O}(1/m^2)\approx r_{\rm g}^2/(224\pi^2\tau^4)$
where $K=12r_{\rm g}^2/\tau^6$, the Kretschmann scalar and $C=\frac{1}{432}r_{\rm g}^2/\tau^6$ 
other contractions of the Riemann tensor.
The trace is usually the most divergent contribution since it includes the inverse metric. By comparing it
with the Planck scale $t_{\rm Pl}$, we are able to determine
$\tau_\star$ where presumably quantum gravity effects become important. Therefore, we demand
$\langle T(\tau_\star)\rangle_{\rm ren}=t_{\rm Pl}^{-2}$. Hence, we 
locate the high-energy scale at $\tau_\star\approx \frac{1}{7} \sqrt{r_{\rm g}t_{\rm Pl}}$. For stellar-mass Schwarzschild black holes, the Zeno border
will then be located $\tau_\star<\tau_\mathcal{Z}^\ell\le r_{\rm g}$ for all initial conditions coming 
from sub-Planckian energy densities. 

Next we consider observables in $\mathcal{Z}_\mathcal{B}$ and introduce 
an auxiliary source functional $\mathcal{J}: \mathcal{C}(\Sigma_t) \rightarrow \mathbb{C}$, 
describing the absorption and emission of an instantaneous field configuration by the 
associated local current density $J:\Sigma_t\rightarrow\mathbb{C}$. Let 
$\Psi_t^{\;J}[\phi] :=\langle\phi|\exp(\mathcal{J}_t)[\Phi]|\Psi_t\rangle$ 
which allows to replace compositions 
of the configuration operator $\Phi$ by the corresponding
succession of functional derivatives with respect to the current. 
In the presence of the auxiliary source, 
\begin{eqnarray}
\label{gse}
	&&\langle\Psi_t^{\; J}|\left|\Phi[f]\right|^2|\Psi_t^{\; J}\rangle
	=
	\nonumber \\
	&&[f]\delta_J^{\; 2}
	\exp\left\{
		\tfrac{1}{4} \tfrac{1}{\sqrt{{\rm det}(g_\Sigma)}}
		[J]\left[{\rm Re}(\mathcal{K}_t)\right]^{-1}[J]
	\right\}
	\|\Psi_t^{\; J}[\phi]\|^2
\end{eqnarray}
with $[f]\delta_J^{\; 2}$ denoting the second functional derivative
with respect to $J$, smeared with an appropriate field configuration $f$.
In the absence of an auxiliary source, the ground-state expectation (\ref{gse})
is real and semi-positive definite. 
It is monotonously decreasing in $\mathcal{Z}_\mathcal{B}$ towards $\Sigma_0$
and vanishes on $\Sigma_0$. 
Similarly, 
\begin{eqnarray}
\label{gsep}
	&&\langle \Psi_t^{\; J}|
	|\Pi|^2[f]
	|\Psi_t^{\; J}\rangle
	= [f]\mathcal{K}_t[f] \; \|\Psi_t^{\; J}[\phi]\|^2
	\nonumber \\
	&&\; -{\rm det}(g_{_\Sigma})k^2(t)
	\langle\Psi_t^{\; J}||\Phi[f]|^2|\Psi_t^{\; J}\rangle_{\Sigma_t}
	\;.
\end{eqnarray}
Both, the real and imaginary part of (\ref{gsep}) are monotonously 
decreasing in $\mathcal{Z}_\mathcal{B}$ towards $\Sigma_0$ and vanish on $\Sigma_0$
when the auxiliary current is switched off. 

Within the Zeno region, the ground-state expectation values \eqref{gse} and \eqref{gsep} 
are controlled by the contractive representation of the evolution semigroup. As a result
$\langle \Psi_t^{\; J}|\mathcal{H}[f]|\Psi_t^{\; J}\rangle$ does receive decreasing
probabilistic support within $\mathcal{Z}_\mathcal{B}$ and has vanishing
probabilistic support on $\Sigma_0$.
The Zeno region completes $\mathcal{B}$ since information cannot leak through 
the geodesic information sink $\Sigma_0$. 
These results remain true if (self-) interactions are included \cite{hof17}
which can be understood at the qualitative level from the approximate 
operator identity 
$\mathcal{H}\cong\mathcal{G}(\Pi^*,\Pi)$. 

\section{Discussion}
The mathematical spacetime model of Schwarzschild black-hole interiors 
is geodesic incomplete. 
This result is often used to argue for the breakdown of general relativity 
as predicted by general relativity. 
In this letter we have shown that interiors of Schwarzschild black holes 
contain so-called Zeno regions, i.e.~stacks of hypersurfaces in the vicinity 
of the geodesic spacetime singularity, on which observables enjoy a
vanishing probabilistic support towards the geodesic spacetime border.
In this sense Zeno regions represent a probabilistic completion of 
Schwarzschild black holes within the usual semiclassical framework. 
Moreover, geodesic incompleteness of Schwarzschild black holes 
qualifies the mathematical spacetime model, but certainly not 
the physical spacetime with a Zeno region in the immediate vicinity
of the geodesic singularity. This statement is justified since 
Zeno regions allow for a consistent evolution of quantum fields 
but are beyond the scope of classical point particle mechanics.
It is interesting to speculate whether Zeno regions are 
terminated by Hawking's hidden surface to which the principle of ignorance
applies \cite{hawk76}, but this we leave
for further investigations.

\acknowledgements
We thank Kerstin Paech for naming the new border after the 
Greek philosopher Zeno of Elea. Thanks to 
Maximilian K\"ogler, Robert Myers, and Florian Niedermann for 
delightful discussions and their thoughts on the topic.
We appreciate financial support from the Alexander von Humboldt-Stiftung.

\bibliography{ssccl}
\end{document}